\documentclass[12pt]{article}

\usepackage{amsfonts,amssymb,amsmath,amscd,amsthm}
\usepackage[all,cmtip]{xy}

\usepackage{epsfig,ulem}

\newcommand{\ra}{\rightarrow}

\newcommand{\ZZ}{{\mathbb Z}}

\newcommand{\RR}{{\mathbb R}}
\newcommand{\CC}{{\mathbb C}}

\newcommand{\cH}{{\mathcal H}}

\newcommand{\eps}{\epsilon}

\newcommand{\cA}{{\mathcal A}}

\newcommand{\Hom}{{\rm Hom}}

\newcommand{\sq}{{\mathsf q}}

\newcommand{\Vect}{{\mathsf{Vect}}}
\renewcommand{\L}{\lambda}

\newtheorem*{prop}{Proposition}

\newcommand{\cD}{{\mathcal D}}

\newcommand{\End}{{\rm End}}

\newcommand{\sA}{{\mathsf A}}

\renewcommand{\sb}{{\mathsf g}}

\newcommand{\ord}{{\rm ord}}
\newcommand{\lcm}{{\rm lcm}}

\renewcommand{\AA}{{\mathsf A}}

\def\be{\begin{equation}}
\def\ee{\end{equation}}
\def\bear{\begin{eqnarray}}
\def\eear{\end{eqnarray}}

\def\eps{{\epsilon}}

\def\cT{{{\mathcal T}}}

\def\cC{{{\mathcal C}}}
\def\cL{{{\mathcal L}}}

\title{Surface operators in 3d Topological Field Theory and 2d Rational Conformal Field Theory}

\author{Anton Kapustin\\{\small \it California Institute of Technology, Pasadena, CA 91125,
U.S.A.}\\
Natalia Saulina\\{\small\it Perimeter Institute, Waterloo, Canada}}

\begin{document}

\begin{titlepage}

\maketitle

\begin{abstract}

We study surface operators in 3d Topological Field Theory and their relations with 2d Rational Conformal Field Theory. We show that a surface operator gives rise to a consistent gluing of chiral and anti-chiral sectors in the 2d RCFT. The algebraic properties of the resulting 2d RCFT, such as the classification of symmetry-preserving boundary conditions, are expressed in terms of properties of the surface operator. We show that to every surface operator one may attach a Morita-equivalence class of symmetric Frobenius algebras in the ribbon category of bulk line operators. This provides a simple interpretation of the results of Fuchs, Runkel and Schweigert on the construction of 2d RCFTs from Frobenius algebras. We also show that every topological boundary condition in a 3d TFT gives rise to a commutative Frobenius algebra in the category of bulk line operators. We illustrate these general considerations  by studying in detail surface operators in abelian Chern-Simons theory.

\end{abstract}

\vspace{-6in}
\parbox{\linewidth}
{\small\hfill \shortstack{pi-strings-195}} \vspace{6in}

\end{titlepage}

\section{Introduction}

There is a well-known relationship between 2d Rational Conformal Field Theory (RCFT) and 3d Topological Field Theory (TFT). Formally, this relationship arises from the fact that both conformal blocks of a chiral algebra of a 2d RCFT and bulk line operators in a semi-simple 3d TFT form a modular tensor category (i.e. a semi-simple ribbon category satisfying a certain non-degeneracy condition). More physically, one may consider a 3d TFT on a cylinder $D^2\times\RR$ and find \cite{Witten, EMSS} that  this is equivalent to a chiral sector of a 2d RCFT living on the boundary $S^1\times\RR$. This may be regarded as a toy model for holographic duality.

The simplest example of this relationship is probably the $U(1)$ Chern-Simons theory at level $k\in 2\ZZ$ which is holographically dual to a free boson RCFT with radius $R^2=k$. A more complicated example is the duality between Chern-Simons gauge theory with a simple Lie group $G$ and the Wess-Zumino-Witten model in 2d.

There is more to a 3d TFT than bulk line operators. For example, we may consider surface operators which are localized on 2d submanifolds. Surface operators form a 2-category whose 1-morphisms are line operators sitting at the junction of two surface operators, and whose 2-morphisms are local operators sitting at the junction of two line operators \cite{ETFT}. The category of bulk line operators can be recovered from this 2-category: it is the category of 1-morphisms from the ``invisible'' surface operator to itself. The ``invisible'' surface operator is equivalent to no surface operator at all; more formally one may describe it by saying that the 2-category of surface operators has a monoidal structure given by fusing surface operators together, and the ``invisible'' surface operator is the identity object with respect to this monoidal structure. 

In this note we discuss the following question: what is the meaning of surface operators in a 3d TFT from the point of view of the corresponding 2d RCFT? We will argue that surface operators in a given 3d TFT correspond to consistent gluings of  chiral and anti-chiral sectors to form a modular-invariant 2d RCFT. More precisely, this interpretation applies to surface operators which do not admit nontrivial local operators. General surface operators correspond to consistent ways of gluing chiral and anti-chiral sectors to a 2d TFT. We illustrate the correspondence using the example of $U(1)_{k}$ current algebra. 

One motivation for this work was the desire to give a more intuitive interpretation to some of the results of Fuchs, Runkel, Schweigert  and collaborators on the algebraic construction of 2d RCFTs from modular tensor categories (see \cite{FRS,FRS2,FRS3,FRS4,FRS5, FFRS} and references therein; an overview of this series of papers is contained  in \cite{FRS0}). These authors related 2d RCFTs with a fixed chiral algebra with Morita-equivalence classes of special symmetric Frobenius algebras in the corresponding modular tensor category. We will see that this results has a very natural explanation in terms of surface operators. A similar argument shows how to associate a commutative  Frobenius algebra in the modular tensor category to any topological boundary condition in a 3d TFT.  This can be regarded as a 3d analogue of the boundary-bulk map in 2d TFT. We describe this map in the case of $U(1)_k\times U(1)_{-k}$ Chern-Simons theory which admits many topological boundary conditions.

The authors are grateful to I. Runkel for comments on a preliminary draft, to A. Tsymbalyuk for the proof of the proposition presented in the appendix, and to Shlomit Wolf for drawing the more elaborate figures.  A. K. would like to thank the Institute for Advanced Studies, Hebrew University, for hospitality during the completion of this work. The work of A. K. was supported in part by the DOE grant DE-FG02-92ER-40701.

\section{Generalities}

\subsection{Surface operators and consistent gluings}

Consider a 3d TFT $\cT$ corresponding to a given chiral algebra $\cA$. This means that the space of conformal blocks of $\cA$ on a Riemann surface $\Sigma$ is the space of states of the 3d TFT on $\Sigma$. A good example to keep in mind is  Chern-Simons theory with gauge group $G$, in which case $\cA$ is the affine Lie algebra corresponding to $G$. We will denote by $\cH_\Sigma$ the space of conformal blocks on $\Sigma$; it is acted upon by the mapping class group of $\Sigma$. When we consider both chiral and anti-chiral sectors, the relevant representation of the mapping class group is $\cH_\Sigma\otimes\cH_\Sigma^*$. To obtain a well-defined 2d RCFT one needs to pick an invariant vector in all representation spaces $\cH_\Sigma\otimes\cH_\Sigma^*$.This choice must be compatible with cutting $\Sigma$ into pieces, so that Segal's axioms \cite{segalCFTaxioms} are satisfied. 

From the 3d viewpoint the space $\cH_\Sigma\otimes\cH_\Sigma^*$ is the space of states of the 3d TFT $\cT\otimes \bar\cT$, where the bar denotes parity-reversal. There is a very natural way to get an invariant vector in this space. Consider the 3-manifold $\Sigma\times [0,1]$. Its boundary is a disjoint union of two copies of $\Sigma$. Let us pick a topological boundary condition in the theory $\cT\otimes\bar\cT$ and impose it on $\Sigma\times\{1\} $. The other copy of $\Sigma$ is a ``cut'' boundary on which we do not impose boundary conditions. 3d TFT assigns to such a 3-manifold a diffeomorphism-invariant vector in $\cH_\Sigma\otimes\cH_\Sigma^*$. Compatibility with cutting and pasting is clearly ensured by the locality of the topological boundary condition imposed at $\Sigma\times \{1\}$.

By the usual folding trick we may reinterpret the boundary condition for the theory $\cT\otimes\bar\cT$ as a surface operator in the theory $\cT$. Thus we consider the theory $\cT$ on $\Sigma\times [0,2]$ with an insertion of a surface operator at $\Sigma\times \{1\}$. Chiral and anti-chiral degrees of freedom then live at $\Sigma\times \{0\}$ and $\Sigma\times\{2\}$, and the surface operator glues them together into a consistent 2d RCFT.

Among all surface operators there is always a trivial one equivalent to no surface operator at all. It corresponds to the diagonal gluing, and the corresponding vector in $\cH_\Sigma\otimes\cH_\Sigma^*=\End \, \cH_\Sigma$ is ${\bf 1}$. From an abstract viewpoint, surface operators in a 3d TFT form a monoidal 2-category (i.e. a 2-category with an associative but not necessarily commutative tensor product) \cite{ETFT}, and the trivial surface operator is the identity object with respect to the product. 

Given this correspondence between surface operators and consistent gluings we may interpret various properties of 2d RCFT in 3d terms. Recall that the main object of study in 2d RCFT is the modular tensor category $\cC$ which from the 3d viewpoint is the category of bulk line operators in the theory $\cT$. For example, if  $\cT$ is Chern-Simons theory, bulk line operators are Wilson lines labeled by integrable HW representations of the corresponding affine Lie algebra. Primary vertex operators in a 2d RCFT are labeled by a pair of objects $W$ and $W'$ of $\cC$. From the 3d viewpoint a primary $V_{WW'}$ inserted at $p\in\Sigma$ is represented by a pair of bulk line operator inserted at $p\times [0,1]$ and $p\times [1,2]$ and labeled by $W$ and $W'$ respectively. These two bulk line operators meet at the point $p\times\{1\}$ which lies on the surface operator. Thus primaries labeled by $W,W'$ are in 1-1 correspondence with local operators living on the surface operator which ``convert'' the bulk line operator $W$ into the bulk line operator $W'$. The space of such local operators is the vector space which 3d TFT attaches to $S^2$ with the surface operator inserted at the equator and with $W$ and $W'$ inserted at the north and south poles (see Fig. 1).

\begin{figure}[htbp] \label{fig1}
\begin{picture}(10,10)(-232,0)
\put(0,0){$W$}
\end{picture}
\begin{center}
\includegraphics{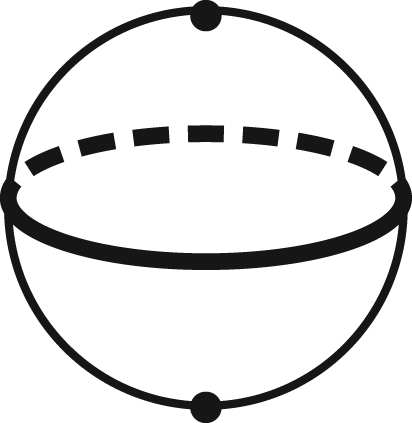}
\end{center}
\begin{picture}(10,10)(-232,-5)
\put(0,0){$W'$}
\end{picture}

\caption{The space of RCFT primaries labeled by objects $W,W'$ is the space of states of the 3d TFT on $S^2$ with insertions of line operators at the poles and the surface operator at the equator.}
\end{figure}
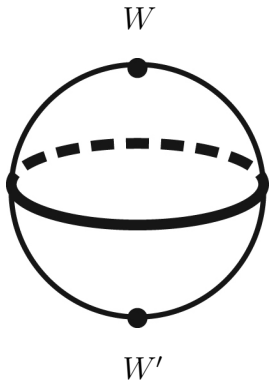

If we take both $W$ and $W'$ to be trivial line operators (i.e. identity objects in the monoidal category $\cC$), then we conclude that the space of primaries transforming in the vacuum representation of the chiral-anti-chiral algebra $\cA\otimes\bar \cA$ is isomorphic to the space of local operators on the surface operator (i.e. the space attached to $S^2$ with an insertion of the surface operator at the equator). One usually requires this space to be one-dimensional (this requirement is called uniqueness of the vacuum). To satisfy this axiom, the surface operator used for gluing should not admit any nontrivial local operators. 

Another interesting question is the classification of boundary conditions for a 2d RCFT.  For concreteness, let $\Sigma$ be a disc. By definition, the {\it double} of such $\Sigma$ is a pair of disks $D^2_+$ and $D^2_-$ glued along their boundaries.  We will identify the double with a unit sphere in $\RR^3$, so that $D^2_+$ and $D^2_-$ correspond to southern and northern hemispheres respectively.  We wish to consider the 3d TFT on the  ball $D^3$ which is the interior of the unit sphere and with an insertion of a surface operator at the equatorial plane (see Fig. 2). Chiral and anti-chiral degrees of freedom live on $D^2_+$ and $D^2_-$ respectively. The support of the surface operator is a disk, so in order to specify the physics completely we need to specify how to terminate the surface operator. That is, we need to pick a line operator $\lambda$ which sits at the boundary of the surface operator $S$. Such line operators can be thought of as 1-morphisms in the monoidal 2-category of surface operators (specifically, morphisms from the trivial surface operator ${\bf 1}$ to the surface operator $S$). We conclude that boundary conditions for the 2d RCFT are labeled by such 1-morphisms.

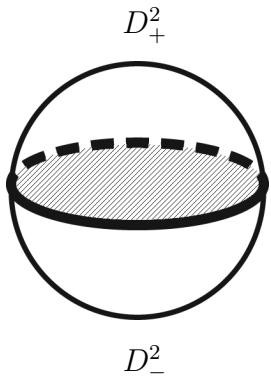
\begin{figure}[htbp] \label{fig2}
\begin{picture}(10,10)(-232,0)
\put(0,0){$D^2_+$}
\end{picture}
\begin{center}
\includegraphics{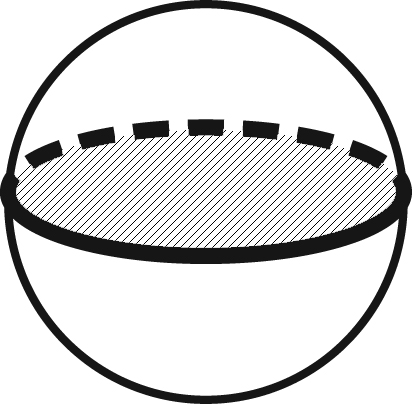}
\end{center}
\begin{picture}(10,10)(-232,-5)
\put(0,0){$D^2_-$}
\end{picture}

\caption{3d TFT on a ball with an insertion of a surface operator (shaded region) on the equatorial plane. Chiral and anti-chiral sectors of the RCFT live on northern and southern hemispheres respectively. The surface operator terminates on a line operator.}
\end{figure}

In the special case $S={\bf 1}$ such line operators are the same as bulk line operators. Thus we end up with a familiar result that, in the case of diagonal gluing, boundary conditions preserving the full chiral algebra $\cA$ are in 1-1 correspondence with representations of $\cA$.

Boundary-changing vertex operators can also be described in the language of surface operators. Let $\L$ and $\L'$ be two line operators on which a given surface operator $S$ may terminate. As mentioned above they can be regarded as 1-morphisms from the trivial surface operator ${\bf 1}$ to the surface operator $S$.  Now, since we are dealing with a 2-category, we may consider morphisms between 1-morphisms, i.e. 2-morphisms. They form a vector space (the 2-category of surface operators is $\CC$-linear) which from the viewpoint of the 3d TFT is the vector space attached to the picture shown in Fig. 3. In other words, this is the space of local operators which can be inserted at the junction of $\L$ and $\L'$. Elements of this space label boundary-changing vertex operators which transform in the vacuum representation of the chiral algebra. In the special case of diagonal gluing where $S={\bf 1}$ and $\L$ and $\L'$ are bulk line operators, this is simply the space of morphisms between objects $\L$ and $\L'$ in the category $\cC$. 

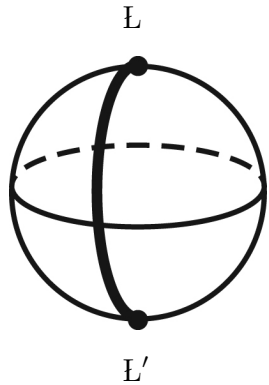
\begin{figure}[htbp] \label{fig3}
\begin{picture}(10,10)(-232,0)
\put(0,0){$\L$}
\end{picture}
\begin{center}
\includegraphics{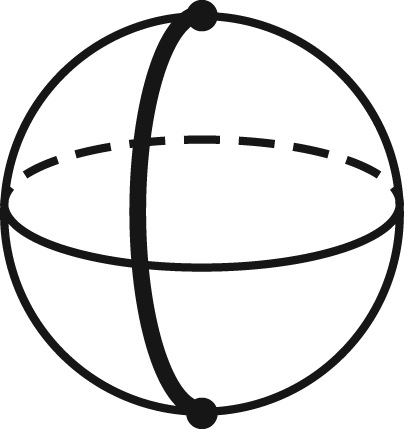}
\end{center}
\begin{picture}(10,10)(-232,-5)
\put(0,0){$\L'$}
\end{picture}

\caption{Boundary conditions are labeled by line operators on which the surface operator can terminate. Given two such line operators $\L,\L'$, the space of boundary-changing operators in the vacuum representation of the chiral algebra is the space of states of the 3d TFT on $S^2$ with insertions of a surface operator (thick line) along a meridian semicircle and line operators $\L$ and $\L'$ at the poles.}
\end{figure}

\subsection{From surface operators to algebra-objects}

Now let us make contact with the work of Fuchs, Runkel and Schweigert \cite{FRS}. Let $S$ be a surface operator and let us pick a line operator $\L$ on which $S$ can terminate (i.e. a 1-morphism from the trivial surface operator ${\bf 1}$ to the surface operator $S$). Let $\bar \L$ be the line operator obtained from $\L$ by orientation reversal; it can be regarded as a 1-morphism from $S$ to the trivial surface operator. Consider now a long rectangular strip of $S$ bounded on the two sides by $\L$ and $\bar \L$ (see Fig. 4).  Since we are dealing with a 3d TFT, the strip can be regarded as arbitrarily narrow and so is equivalent to a bulk line operator which we will denote $W(S,\L)$. More formally, if we regard $\L$ as a 1-morphism from ${\bf 1}$ to $S$ and ${\bar \L}$ as a 1-morphism from $S$ to ${\bf 1}$, then their composition $W(S,\L)=\bar \L\circ \L$ is a 1-morphism from ${\bf 1}$ to ${\bf 1}$ and so is an object of the modular tensor category $\cC$.

\begin{figure}[htbp] \label{fig4}
\begin{picture}(10,10)(-195,50)
\put(0,0){$\mathbf\L$}
\end{picture}
\begin{center}
\includegraphics{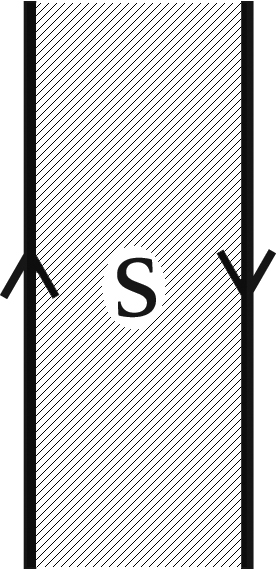}
\end{center}

\caption{A strip of a surface operator $S$ bounded from the left by a line operator $\L$ and from the right by its orientation-reversal.}
\end{figure}
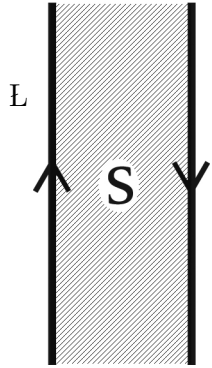

The object $W(S,\L)$ is very special: it is a symmetric Frobenius algebra in the modular tensor category $\cC$. Let us remind what this means. First of all, an algebra $\AA$ in a monoidal category $\cC$ is an object $\AA$ equipped with a morphism $m:\AA\otimes\AA\ra\AA$ (product) satisfying the associativity constraint expressed as the commutativity of the following diagram:

\vspace{0.2cm}

\xymatrixcolsep{4pc}
\centerline{ \xymatrix{
(\AA\otimes \AA)\otimes \AA \ar[r]^{m\otimes id}\ar[dd]^a &\AA\otimes \AA \ar[rd]^m & \\
& & \AA\\
\AA \otimes (\AA \otimes \AA)\ar[r]^{id\otimes m}& \AA\otimes \AA \ar[ru]^m & }}

\vspace{0.2cm}

Instead of drawing commutative diagrams, it is convenient to use the graphical calculus for monoidal categories explained for example in \cite{FRS} and \cite{BK}. An object $A$ of a monoidal category is then represented by an upward going line labeled by $A$, the tensor product  of two objects is represented by drawing two lines going in the same direction, and a morphism  from $A_1\otimes\ldots \otimes A_n$ to $B_1\otimes\ldots \otimes B_k$ is represented by a vertex with $n$ incoming and $k$ outgoing lines. Composition of morphisms corresponds to the concatenation of the diagrams in the vertical direction. The vertex corresponding to the identity morphism is usually omitted, so that we have
\vspace{0.2cm}

\setlength{\unitlength}{1cm}
\centerline{
\begin{picture}(2,2)
\put(0,1){
\put(0,0){$id_A=$}\put(1.5,-0.5){\line(0,1){1}}\put(1.3,-1){$A$}\put(1.3,0.7){$A$}
}
\end{picture}
}

Then the associativity constraint on $m$ is expressed at the equality of morphisms corresponding to the two diagrams shown below.

\vspace{0.2cm}

\setlength{\unitlength}{1cm}
\centerline{
\begin{picture}(2,2)
  \linethickness{0.25mm} \qbezier(0,0)(0.01,0.33)(0.3,0.5)\qbezier(0.3,0.5)(0.59,0.33)(0.6,0)
\put(0.3,0.51){\circle*{0.1}}\put(-0.17,-0.4){$\AA$}\qbezier(0.3,0.5)(0.31,0.83)(0.6,1)
\put(0.5,-0.4){$\AA$}\put(0.6,1){\circle*{0.1}}\put(0.6,1){\line(0,1){0.5}}\qbezier(0.6,1)(1.29,0.33)(1.3,0)
\put(1.2,-0.4){$\AA$}\put(0.5,1.6){$\AA$}\put(1.7,0.5){$\mathbf =$}
\qbezier(4,0)(4.01,0.33)(4.3,0.5)\qbezier(4.3,0.5)(4.59,0.33)(4.6,0)\put(4.3,0.51){\circle*{0.1}}
\qbezier(4.3,0.5)(4.29,0.83)(3.9,1)\put(3.9,1){\circle*{0.1}}\put(3.9,1){\line(0,1){0.5}}
\qbezier(3.2,0)(3.21,0.33)(3.9,1)\put(3,-0.4){$\AA$}\put(3.8,-0.4){$\AA$}
\put(4.5,-0.4){$\AA$}
\put(3.8,1.6){$\AA$}
\put(-0.2,0.5){$m$}\put(0.1,1){$m$}\put(4.4,0.5){$m$}\put(4,1){$m$}
\end{picture}}

\vspace{0.6cm}

In the  case of $W(S,\L)$ the morphism $m$ arises from the ``open-string vertex'' shown  in Fig. 5. The associativity constraint is satisfied because the two diagrams shown in Fig. 6 can be deformed into one another. Thus $W(S,\L)$ is an algebra in the category of bulk line operators.

Similarly, a coalgebra-object is an object $\AA$ of $\cC$ equipped with a morphism $\Delta:\AA\ra \AA\otimes\AA$ (coproduct) satisfying the coassociativity constraint expressed as the commutativity of the following diagram:

\vspace{0.2cm}

\xymatrixcolsep{4pc}
\centerline{ \xymatrix{
 &\AA\otimes \AA \ar[r]^{\Delta \otimes id} & (\AA\otimes \AA)\otimes \AA \ar[dd]^a\\
\AA\ar[ru]^{\Delta}\ar[rd]^{\Delta}& &\\
&\AA\otimes \AA \ar[r]^{id\otimes \Delta}& \AA\otimes(\AA\otimes \AA) }}

\vspace{0.2cm}

In the case of $W(S,\L)$ the coproduct is given by the diagram shown in Fig. 7. The coassociativity follows from the fact that the two diagrams shown in Fig. 8 can be deformed into one another.

\begin{figure}[htbp] \label{fig7}
\begin{center}
\includegraphics{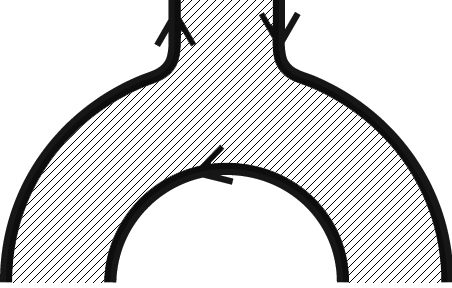}
\end{center}

\caption{The product.}
\end{figure}

\begin{figure}[htbp] \label{fig12a}
\begin{center}
\includegraphics{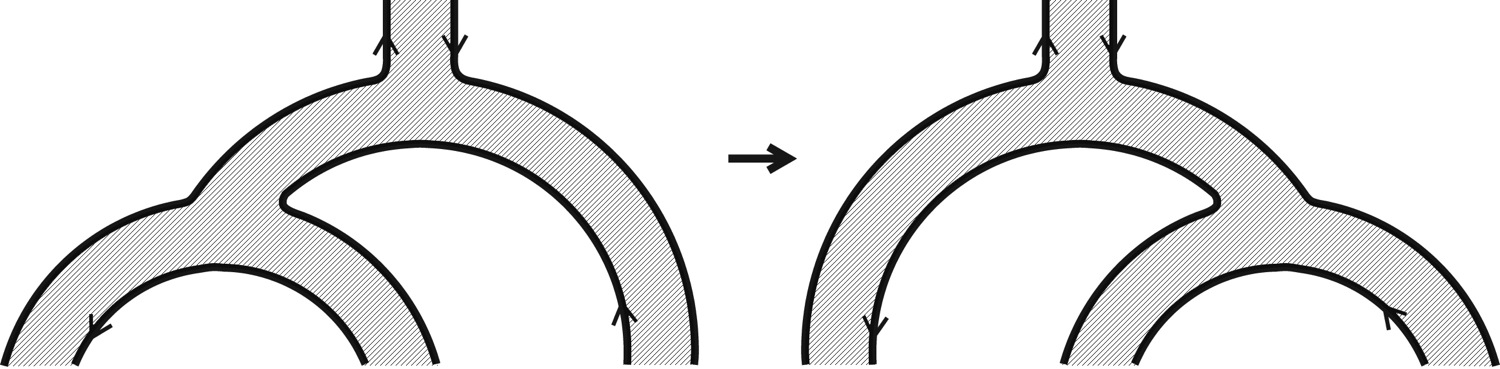}
\end{center}

\caption{The associativity constraint.}
\end{figure}

\begin{figure}[htbp] \label{fig11}
\begin{center}
\includegraphics{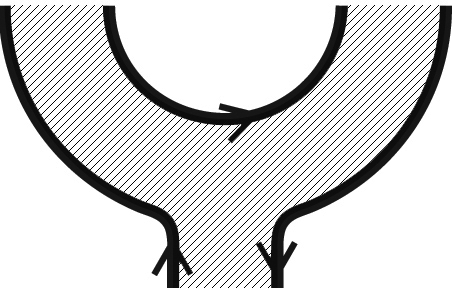}
\end{center}

\caption{The coproduct.}
\end{figure}

\begin{figure}[htbp] \label{fig12}
\begin{center}
\includegraphics{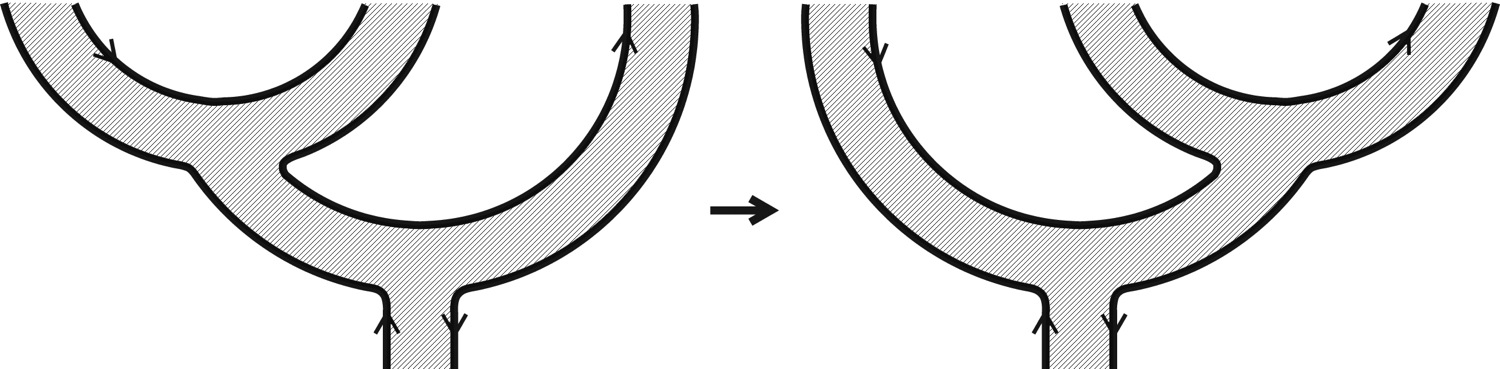}
\end{center}

\caption{The coassociativity constraint.}
\end{figure}

A Frobenius algebra in $\cC$ is an object $\AA$ which is both an algebra and a coalgebra in $\cC$ such that $m$ and $\Delta$ are compatible in the sense that the following equalities hold:
$$(id_{\AA}\otimes m)\circ (\Delta \otimes id_{\AA})=\Delta\circ m=(m\otimes id_{\AA})\circ
(id_{\AA}\otimes \Delta)$$
The equalities can be represented graphically as follows (all vertices are labeled either by $m$ or by $\Delta$):
 
\vspace{0.8cm}
\setlength{\unitlength}{1cm}
\centerline{\begin{picture}(2,2)
\linethickness{0.25mm} \qbezier(0,1)(0.01,0.67)(0.3,0.5)\qbezier(0.3,0.5)(0.59,0.67)(0.6,1)
\put(0.3,0.49){\circle*{0.1}}\put(0.3,0.48){\line(0,-1){0.5}}\put(0,1){\line(0,1){1}}\put(-0.15,2.2){$\AA$}
\put(0.2,-0.45){$\AA $}
\qbezier(0.6,1)(0.61,1.33)(0.9,1.5)\qbezier(0.9,1.5)(1.19,1.33)(1.2,1)
\put(0.9,1.51){\circle*{0.1}}\put(0.9,1.52){\line(0,1){0.5}}\put(0.8,2.1){$ \AA$}
\put(1.2,1){\line(0,-1){1}}\put(1.1,-0.45){$\AA$}\put(2,1.2){$=$}
\end{picture} \hspace{0.2cm}
\begin{picture}(2,2)
\linethickness{0.25mm} \qbezier(0.6,2)(0.61,1.67)(0.9,1.5)
\qbezier(0.9,1.5)(1.19,1.67)(1.2,2)
\put(0.9,1.49){\circle*{0.1}}\put(0.9,1.48){\line(0,-1){0.5}}
\put(1.1,2.15){$\AA $}\put(0.5,2.15){$\AA $}
\qbezier(0.6,0)(0.61,0.33)(0.9,0.5)\qbezier(0.9,0.5)(1.19,0.33)(1.2,0)
\put(0.9,0.51){\circle*{0.1}}\put(0.9,0.52){\line(0,1){0.5}}
\put(0.45,-0.4){$ \AA$}\put(1.1,-0.4){$ \AA$}
\end{picture} \hspace{0.2cm}
\begin{picture}(2,2)
\linethickness{0.25mm} \qbezier(1.2,1)(1.19,0.67)(0.9,0.5)\qbezier(0.9,0.5)(0.61,0.67)(0.6,1)
\put(0.9,0.49){\circle*{0.1}}\put(0.9,0.48){\line(0,-1){0.5}}\put(1.2,1){\line(0,1){1}}
\put(1.2,2.2){$\AA$}
\put(0.8,-0.45){$\AA $}
\qbezier(0.6,1)(0.59,1.33)(0.3,1.5)\qbezier(0.3,1.5)(0.01,1.33)(0,1)
\put(0.3,1.51){\circle*{0.1}}\put(0.3,1.52){\line(0,1){0.5}}\put(0.2,2.2){$ \AA$}
\put(0,1){\line(0,-1){1}}\put(-0.1,-0.45){$\AA$}\put(-0.8,1.2){$=$}
\end{picture}}

\vspace{0.6cm}

In the case of $W(S,\L)$ the compatibility of $m$ and $\Delta$ follows from the fact that the three diagrams shown in Fig. 9 can be deformed into one another.

\begin{figure}[htbp] \label{fig14}
\begin{center}
\includegraphics{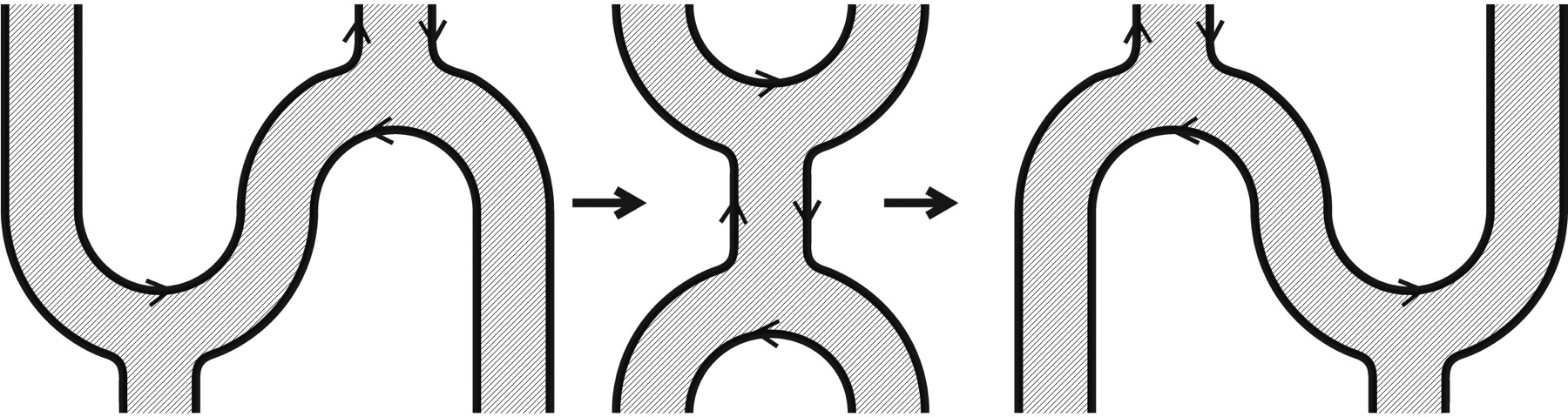}
\end{center}

\caption{The Frobenius condition.}
\end{figure}

An algebra $\AA$ in a monoidal category $\cC$ is equipped with a unit if one is given a morphism $\iota_\AA:{\bf 1}\ra\AA$ such that the following three pictures correspond to the same morphism:

\vspace{0.2cm}

\setlength{\unitlength}{1cm}
\centerline{\begin{picture}(2,2)
\linethickness{0.25mm} 
\put(-2.5,0){\line(0,1){1}}\put(-2.5,-0.05){\circle{0.15}}\put(-3.8,0.5){$\iota_{\AA}=$}
\qbezier(0,0)(0.01,0.33)(0.3,0.5)\qbezier(0.3,0.5)(0.59,0.33)(0.6,0)
\put(0.3,0.51){\circle*{0.1}}\put(0.3,0.52){\line(0,1){0.5}}\put(0,-0.08){\circle{0.15}}
\put(0.52,-0.4){$\AA$}\put(0.22,1.14){$\AA $}\put(0.6,0.5){$=$}
\put(1.5,0){\line(0,1){1}}\put(1.4,-0.4){$\AA $}\put(1.4,1.2){$\AA $}\put(1.9,0.5){$=$}
\end{picture}\hspace{0.3cm}
\begin{picture}(2,2)
\linethickness{0.25mm}
\qbezier(0,0)(0.01,0.33)(0.3,0.5)\qbezier(0.3,0.5)(0.59,0.33)(0.6,0)
\put(0.3,0.51){\circle*{0.1}}\put(0.3,0.52){\line(0,1){0.5}}\put(0.58,-0.1){\circle{0.15}}
\put(-0.1,-0.4){$\AA$}\put(0.22,1.14){$\AA $}
\end{picture}}
\vspace{0.6cm}

A coalgebra is equipped with a counit if one is given a morphism $\eps_\AA:\AA\ra {\bf 1}$ such that the the following three pictures correspond to the same 
morphism:
 
 \vspace{0.2cm}

\setlength{\unitlength}{1cm}
\centerline{\begin{picture}(2,2)
\linethickness{0.25mm} 
\put(-2.5,0){\line(0,1){1}}\put(-2.5,1.05){\circle{0.15}}\put(-3.6,0.4){$\eps_{\AA}=$}
\qbezier(0,1)(0.01,0.67)(0.3,0.5)\qbezier(0.3,0.5)(0.59,0.67)(0.6,1)
\put(0.3,0.49){\circle*{0.1}}\put(0.3,0.49){\line(0,-1){0.5}}\put(0,1.08){\circle{0.15}}
\put(0.52,1.2){$\AA$}\put(0.21,-0.38){$\AA $}
\put(0.8,0.5){$=$}
\put(1.5,0){\line(0,1){1}}\put(1.4,-0.4){$\AA $}\put(1.4,1.2){$\AA $}\put(1.9,0.5){$=$}
\end{picture}\hspace{0.3cm}
\begin{picture}(2,2)
\linethickness{0.25mm} 
\qbezier(0,1)(0.01,0.67)(0.3,0.5)\qbezier(0.3,0.5)(0.59,0.67)(0.6,1)
\put(0.3,0.49){\circle*{0.1}}\put(0.3,0.49){\line(0,-1){0.5}}\put(-0.15,1.2){$\AA$}
\put(0.59,1.08){\circle{0.15}}\put(0.21,-0.38){$\AA $}
\end{picture}}

\vspace{0.6cm}
 
In the case of $W(S,\L)$ the unit and the counit are defined in the obvious way, see Fig. 10.

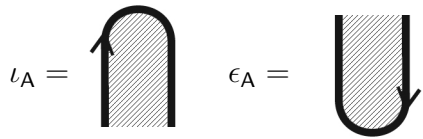
\begin{figure}[htbp] \label{fig17}
\begin{center}
\begin{picture}(1,1) \put(0,0.8){$\iota_\AA=$} \end{picture}\includegraphics{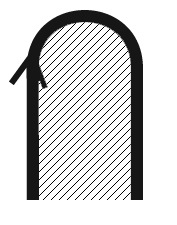}\quad \begin{picture}(1,1) \put(0,0.8){$\eps_\AA=$} \end{picture} \includegraphics{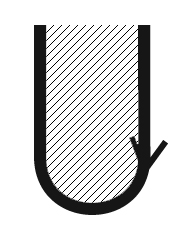} 
\end{center}

\caption{The Frobenius condition.}
\end{figure}

Recall that our category $\cC$ is not only monoidal, but also rigid (because it is a ribbon category).  In a rigid category every object $A$ has a left-dual object ${}^*A$ and a right-dual object $A^*$; these objects are automatically unique up to isomorphism. In a ribbon category the left-dual is automatically isomorphic to the right-dual; a rigid monoidal category with this property is called sovereign. 

From the physical viewpoint, the dual object is obtained by reversing the orientation of the bulk line operator. In the diagrammatic notation this is accounted for by placing arrows on all  lines, so that a line labeled by $A$ but with an arrow pointing down represents the object $A^*={}^*A$. By the definition of a rigid category, for every $A$ we have four morphisms $A\otimes A^*\ra {\bf 1},$ ${\bf 1}\ra A^*\otimes A$, $A^*\otimes A\ra {\bf 1}$ and ${\bf 1}\ra A\otimes A^*$ which are represented by the following four pictures:

 \vspace{0.2cm}

\setlength{\unitlength}{1cm}
\begin{center}
\begin{picture}(2,2)(2,0)
\linethickness{0.25mm} 

\put(0,1){\oval(1,1)[t]}\put(-0.5,0.5){\vector(0,1){0.5}}\put(0.5,1){\line(0,-1){0.5}}\put(0.5,0.8){\vector(0,-1){0}}\put(-0.7,0){$A$}\put(0.3,0){$A$}

\put(2,0){
\put(0,1){\oval(1,1)[b]}\put(-0.5,1.5){\vector(0,-1){0.5}}\put(0.5,1.5){\line(0,-1){0.5}}\put(0.5,1.2){\vector(0,1){0}}\put(-0.7,1.7){$A$}\put(0.3,1.7){$A$}
}

\put(4,0){
\put(0,1){\oval(1,1)[t]}\put(-0.5,1){\line(0,-1){0.5}}\put(-0.5,0.8){\vector(0,-1){0}}\put(0.5,0.5){\vector(0,1){0.5}}\put(-0.7,0){$A$}\put(0.3,0){$A$}
}

\put(6,0){
\put(0,1){\oval(1,1)[b]}\put(-0.5,1){\line(0,1){0.5}}\put(-0.5,1.2){\vector(0,1){0}}\put(0.5,1.5){\vector(0,-1){0.5}}\put(-0.7,1.7){$A$}\put(0.3,1.7){$A$}
}

\end{picture}
\end{center}

\vspace{0.4cm}

These four morphisms satisfy four obvious identities the first of which looks as follows:

 \vspace{0.2cm}

\setlength{\unitlength}{1cm}
\begin{center}
\begin{picture}(4,4)(2,-1.5)
\linethickness{0.25mm} 
\put(0,1){\oval(1,1)[t]}\put(-0.5,-0.5){\line(0,1){1.5}}\put(-0.5,0.5){\vector(0,1){0}}\put(0.5,1){\line(0,-1){0.5}}\put(0.5,0){\vector(0,-1){0}}\put(-0.7,-1){$A$}

\put(1,-1){
\put(0,1){\oval(1,1)[b]}\put(-0.5,1.5){\line(0,-1){0.5}}\put(0.5,1){\line(0,1){1.5}}\put(0.5,1.5){\vector(0,1){0}}\put(0.3,2.7){$A$}
}

\put(2.3,0.5){$=$}\put(3.5,-0.5){\line(0,1){2}}\put(3.5,0.5){\vector(0,1){0}}\put(3.3,-1){$A$}\put(3.3,1.7){$A$}

\end{picture}
\end{center}

It is clear from Fig. 4  that  the object $\AA=W(S,\L)$ is self-dual: ${}^*\AA=\AA^*=\AA$. This means that in all diagrams which involve only $\AA$ we may drop the arrows on lines. Accordingly, two of the four duality morphisms become morphisms from $\AA\otimes\AA$ to ${\bf 1}$, while the other two become morphisms from ${\bf 1}$ to $\AA\otimes\AA$.  It is easy to see that the former two are both given by $\eps_\AA\circ m: \AA\otimes \AA\ra {\bf 1}$. Pictorially this is expressed as follows:
\vspace{0cm}

\setlength{\unitlength}{1cm}
\begin{center}
\begin{picture}(3,2)(0,-0.2)

\put(0,1){\oval(1,1)[t]}\put(-0.5,0.5){\line(0,1){0.5}}\put(0.5,1){\line(0,-1){0.5}}\put(-0.7,0){$\AA$}\put(0.3,0){$\AA$}
\put(1,0.5){$=$}
\put(2,0.5){
\qbezier(0,0)(0.01,0.33)(0.3,0.5)\qbezier(0.3,0.5)(0.59,0.33)(0.6,0)\put(0.3,0.51){\circle*{0.1}}\put(0.3,0.52){\line(0,1){0.5}}\put(0.3,1.06){\circle{0.15}}
\put(-0.2,-0.5){$\AA$}\put(0.4,-0.5){$\AA$}
}

\end{picture}

\end{center}
Similarly, the remaining two duality morphisms are both given by $\Delta\circ \iota_\AA:{\bf 1}\ra \AA\otimes\AA$. 

According to \cite{FRS}, a counital Frobenius algebra $\AA$ in a sovereign monoidal category is called symmetric if the two morphisms from $\AA$ to $\AA^*$ shown below coincide:
 \vspace{0.2cm}

\setlength{\unitlength}{1cm}
\begin{center}
\begin{picture}(4,2)(0,-0.2)
\put(0,0.5){
\qbezier(0,0)(0.01,0.33)(0.3,0.5)\qbezier(0.3,0.5)(0.59,0.33)(0.6,0)\put(0.3,0.51){\circle*{0.1}}\put(0.3,0.52){\line(0,1){0.4}}\put(0.3,0.96){\circle{0.15}}
\put(0,0){\line(0,-1){0.5}}\put(1.1,0){\oval(1,1)[b]}\put(1.6,0){\line(0,1){1.2}}
\put(0,0){\vector(0,1){0}}
\put(0.6,0){\vector(0,1){0}}
\put(1.6,0){\vector(0,-1){0}}
\put(-0.2,-1){$\AA$}
\put(1.4,1.3){$\AA$}
}
\put(2.2,0.5){$=$}
\put(4,0.5){
\qbezier(0,0)(0.01,0.33)(0.3,0.5)\qbezier(0.3,0.5)(0.59,0.33)(0.6,0)\put(0.3,0.51){\circle*{0.1}}\put(0.3,0.52){\line(0,1){0.4}}\put(0.3,0.96){\circle{0.15}}
\put(-0.5,0){\oval(1,1)[b]}\put(-1,0){\line(0,1){1.2}}\put(0.6,0){\line(0,-1){0.5}}
\put(-1,0){\vector(0,-1){0}}
\put(0,0){\vector(0,1){0}}
\put(0.6,0){\vector(0,1){0}}
\put(0.4,-1){$\AA$}
\put(-1.2,1.3){$\AA$}
}
\end{picture}

\end{center}
In the case of $W(S,\L)$ it follows from the above discussion that both morphisms are the identity morphism from $\AA$ to $\AA^*=\AA$, so the Frobenius algebra $W(S,\L)$ is symmetric.

We are assuming all along that the identity object of $\cC$ is simple, i.e. $\Hom({\bf 1},{\bf 1})=\CC$. Equivalently, the 3d TFT is assumed not to admit nontrivial local operators. Hence the composition of a unit and a counit of $\AA$ is a number $\beta_1(\AA)\in\CC$. For the algebra-object $W(S,\L)$ this number has the following meaning: inserting a surface operator $S$ shaped as a disk and bounded by the line operator $\L$ is equivalent to mutliplying the correlator by $\beta_1(W(S,\L))$ (see Fig. 11).

Another natural operation is punching a hole in a surface operator $S$, so that the hole is bounded by the line operator $\L$. Such a hole is equivalent to a local operator on on $S$. If one imposes the uniqueness of vacuum axiom, $S$ does not admit any nontrivial local operators, so this local operator must be the identity operator times a number $\beta_2(\AA)\in\CC$.  Then punching a hole bounded by $\L$ is equivalent to multiplying the correlator by $\beta_2(\AA)$ (see Fig. 12). Even if one does not impose uniqueness of vacuum, one may still require that a hole bounded by $\L$ be equivalent to the identity operator times a complex number $\beta_2(\AA)$. 

 A Frobenius algebra $\AA$ is called special \cite{FRS} if this requirement is satisfied and both $\beta_1(\AA)$ and $\beta_2(\AA)$ are nonzero. Note that the numbers $\beta_1(\AA)$ and $\beta_2(\AA)$ for $\AA=W(S,\L)$ are not independent but satisfy
 $$
 \beta_1(\AA)\beta_2(\AA)=\dim \AA,
 $$
 where $\dim \AA$ is the quantum dimension of $\AA$ (the expectation value of the unknot labeled by the object $\AA$). Indeed, the unknot labeled by $W(S,\L)$ can be thought of  as an annulus of $S$ bounded by $\L$. Such an annulus can be thought of as the result of punching a hole in a disk-shaped $S$ bounded by $\L$. Since the disk without a hole contributes $\beta_1(\AA)$ to the correlator and punching a hole multiplies the correlator by $\beta_2(\AA)$, we get the above identity.\footnote{One can show that this identity holds for an arbitrary special symmetric Frobenius algebra $\AA$ \cite{FRS}.}
 
 If $\AA$ is a special Frobenius algebra, one can rescale $\eps$ by $\lambda\in\CC^*$ and $\Delta$ by $\lambda^{-1}$, so that $\beta_1(\AA)\mapsto\lambda\beta_1(\AA)$ and $\beta_2(\AA)\mapsto \lambda^{-1}\beta_2(\AA)$ \cite{FRS,FFRS}. The normalization condition adopted in \cite{FRS,FFRS} is $\beta_1(\AA)=\dim \AA,$ $\beta_2(\AA)=1$, so that punching a hole does not change the correlator. Another natural  normalization condition is $\beta_1(\AA)=\beta_2(\AA)=\sqrt {\dim\AA}$. It has the advantage of being invariant with respect to tensoring the line operator $\L$ with a finite-dimensional vector space of dimension $n$ (this procedure multiplies $\beta_1$ and $\beta_2$ by $n$ and $\dim\AA$ by $n^2$).

\begin{figure}[htbp] \label{fig18}
\begin{center}
\begin{picture}(4.3,2)\put(0,1){$\eps_\AA\circ \iota_\AA=$}\put(2.1,0.6){\line(0,1){1}}\put(2.1,1.65){\circle{0.15}}\put(2.1,0.55){\circle{0.15}}\put(2.5,1){$=\beta_1(\AA)=$}\end{picture} \includegraphics{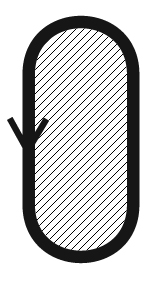}
\end{center}

\caption{Composition of the unit and the counit.}
\end{figure}

\begin{figure}[htbp] \label{fig18ab}
\begin{center}
\begin{picture}(4,3)(1,0)
\includegraphics{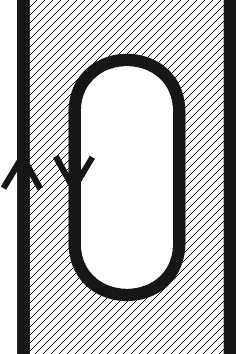}\put(1,1.2){$=\beta_2(\AA)\times$} \put(3,0){\includegraphics{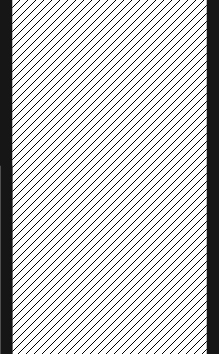}}

\end{picture}
\end{center}

\caption{Punching a hole in a surface operator.}
\end{figure}

We have seen so far that to every surface operator $S$ and a line operator $\L$ on which it can terminate one can attach a symmetric Frobenius algebra in the category $\cC$. On the other hand, the authors of \cite{FRS} show that to any special symmetric Frobenius algebra one can associate a consistent gluing of chiral and anti-chiral sectors. Thus if $S$ is special, we have two ways to associate a consistent gluing to the pair $(S,\L)$: either we ignore $\L$ and simply use the surface operator $S$ to construct such a gluing, or we first compute the algebra $W(S,\L)$ in $\cC$ and them construct the gluing following the recipe of \cite{FRS}. It is not difficult to see that these two procedures give the same result, up to an overall scalar factor. Consider for simplicity the case when the Riemann surface $\Sigma$ has empty boundary. According to the first approach, the gluing is given by a vector $v_S\in\cH_\Sigma\otimes\bar\cH_\Sigma$ which corresponds to the 3-manifold $\Sigma\times I$ with an insertion of the surface operator $S$ at the midpoint of $I$. Using the fact that $S$ is special, we may punch circular holes in $S$ while changing $v_S$ at most by a scalar factor. The surface operator $S$ with holes in it can be deformed into a ribbon graph whose edges are labeled by objects $W(S,\L)$ and vertices correspond to the product morphism $m: W(S,\L)\otimes W(S,\L)\ra W(S,\L)$. Thus we conclude that $v_S$ can be obtained by considering $\Sigma\times I$ with an insertion of a ribbon graph. This is precisely the prescription of \cite{FRS}.

Note that the construction of \cite{FRS} seems to depend not only on the surface operator $S$, but also on the choice of $\L$. The above discussion makes it clear that in fact $v_S$ must be independent of $\L$, up to a scalar factor. This can be demonstrated as follows.  First, we note that given a surface operator $S$ and two line operators $\L$ and $\L'$ on which $S$ can terminate, we have two canonical objects in the category $\cC$: the strip of $S$ bounded from the left by $\L$ and from the right by $\bar\L'$, and the strip of $S$ bounded from the left by $\L'$ and from the right by $\bar\L$. We will denote them $W(S,\L,\L')$ and $W(S,\L',\L)$ respectively.  It is easy to see that these are Morita-equivalence bimodules, so the algebras $W(S,\L)$ and $W(S,\L')$ are Morita-equivalent. Second, it is shown in \cite{FFRS} that up to a scalar factor, the vector $v_S$ depends only on the Morita-equivalence class of the algebra $W(S,\L)$.

\section{Example:  $U(1)$ current algebra}
\subsection{Surface operators in $U(1)$ Chern-Simons theory}
To make our discussion more concrete, in the rest of this paper we will analyze the example of $U(1)$ Chern-Simons theory at level $k=2N$. The corresponding chiral algebra $\cA$ is the $U(1)$ current algebra at level $2N$. We are interested in classifying 2d RCFTs based on this current algebra. The category of conformal blocks (equivalently, the category of Wilson line operators in the $U(1)$ Chern-Simons theory) is a particular example of the following general construction. Let  $\cD$ be a finite abelian group. We will denote by $\Vect_\cD$ the category of $\cD$-graded finite-dimensional complex vector spaces. Its has an obvious monoidal structure given by the ``convolution'': the tensor product of a vector space $E$ in degree $X\in\cD$ and $E'$ in degree $X'\in\cD$ is a vector space $E\otimes E'$ in degree $X+X'$. The monoidal category $\Vect_\cD$ has a deformation parameterized by an element $h\in H^3(\cD,\CC^*)$ which describes the associator morphism. In order for the deformation to be braided monoidal, $h$ must have a special form: it must arise from a quadratic function $\sq:\cD\ra \RR/\ZZ$ \cite{JS}. An explicit formula for $h$ in terms of $\sq$ can be found for example in \cite{Stirling,KS}. The braiding and the ribbon structure are also determined by $\sq$. We will denote by $\Vect_\cD^\sq$ the ribbon category parameterized by the quadratic function $\sq$. 

In case of Chern-Simons theory with $G=U(1)$ at level $k=2N$ we have $\cD=\ZZ/2N\ZZ$; elements of $\cD$ label Wilson line operators modulo those which can terminate on monopole operators (see \cite{KS} for an explanation of this). The quadratic function $\sq:\cD\ra\RR/\ZZ$  has the form
\begin{equation}\label{sq}
\sq(X)=\frac{X^2}{4N},\quad X\in \ZZ/2N\ZZ.
\end{equation}
Simple objects in $\Vect_\cD^\sq$ are labeled by elements of $\cD$; we will denote by $W_X$ the object labeled by $X\in\cD$. The tensor product of simple objects $W_X$ and $W_Y$ is identified with $W_{X+Y}$; the braiding isomorphism can be thought of as a number
$$
c_{X,Y}=\exp(\pi i \sb(X,Y)),
$$
where $\sb(X,Y)$ is the $\RR/\ZZ$-valued bilinear function on $\cD\times\cD$ derived from the quadratic function $\sq$:
$$
\sb(X,Y)=\sq(X+Y)-\sq(X)-\sq(Y)=\frac{XY}{2N}. 
$$
Note that $c_{X,Y}$ is only well-defined up to a sign; to make it completely unambiguous one needs to choose particular representatives in $\ZZ$ for all elements $X\in \ZZ/2N\ZZ$. The physical reason for this is explained in \cite{KS}. Usually one chooses representatives so that they lie in the interval from $0$ to $2N-1$. We denote by $W_i$ the Wilson line operator corresponding to an integer $i$ in this interval.

Surface operators in this theory have been discussed in \cite{KS}. It was shown there that they are 
labelled by divisors of $N,$ i.e. $vm=N$ for some integer $m$. Let us recall how to construct a surface operator corresponding to a divisor $v$. We introduce a periodic scalar $\varphi$ living on the support $\Sigma$ of the surface operator and add the following piece to the Chern-Simons action:
$$\Delta S={i v \over 2\pi}\int_{\Sigma}\varphi (dA^{(+)}-dA^{(-)})$$
where $A^{(\pm)}$ are the limiting values of gauge fields to the left/right of $\Sigma.$ The field $\varphi$ should be viewed as a Lagrange multiplier whose equation of motion enforces the constraint that $v(A^{(+)}-A^{(-)})$ is trivial on $\Sigma$. Let $G^{(\pm)}$ denote the groups of gauge transformations acting on $A^{(\pm)}$. The constraint
$$
v(A^{(+)}-A^{(-)})=0
$$
is invariant under the subgroup of $G^{(+)}\times G^{(-)}$ which consists of elements of the form
$$
\left(\alpha,\alpha+\frac{2\pi\ell}{v}\right),\quad \alpha:\Sigma\ra \RR/2\pi\ZZ,\quad \ell\in\ZZ/v\ZZ.
$$
This means that at the surface $\Sigma$ the gauge group is broken down to $G_0\simeq U(1)\times \ZZ_v$. The functional integral with an insertion of such a surface operator is gauge-invariant and diffeomorphism-invariant if we assign to $\varphi$ charge $w=2m$ with respect to the $U(1)$ factor in $G_0$ and charge $m$ with respect to the $\ZZ_v$ factor in $G_0$ \cite{KS}. 

The ``invisible'' surface operator is a special case corresponding to $v=1$. Indeed, in this case the constraint reads $A^{(+)}=A^{(-)}$, which means that the gauge field is continuous on $\Sigma$.

A surface operator of the above kind may be characterized by the charges of Wilson lines which may meet at a point on the surface operator. That is, imagine a surface operator in $\RR^3$ whose support $\Sigma$ is a hyperplane dividing the space into left and right half-spaces, a Wilson line with charge $X_1$ coming to the surface from the left and a Wilson line with charge $X_2$ coming to the surface from the right, so that Wilson lines meet at $p\in\Sigma$. The net charge $X_1+X_2$ need not be zero because one may insert a local operator of the form $\exp(i\nu\varphi)$. It was shown in section 6.2 of \cite{KS} that for a fixed divisor $v$ $X_1$ and $X_2$ must satisfy
\be \label{lagrv}
X_1-X_2=0 \ \text{mod} \ 2v, \quad  X_1+X_2=0\ \text{mod} \ 2N/v.
\ee
Note that for $v=1$ the second equation says that $X_1=-X_2$ in $\cD$, while the first equation is trivially satisfied. This means that the charge of the Wilson line does not jump at all, which is consistent with the fact that for $v=1$ we are dealing with the ``invisible'' surface operator.

The set of pairs $(X_1,X_2)$ satisfying the above equations defines a subgroup $\cL_v$ of $\ZZ_{2N}\times \ZZ_{2N}$. More generally, a surface operator in any abelian Chern-Simons theory can be described by a subgroup $\cL$ of $\cD\times\cD$ where the finite abelian group $\cD$ classifies charges of bulk Wilson lines \cite{KS}. This subgroup must be Lagrangian with respect to the quadratic form $\sq\oplus (-\sq):\cD\times\cD\ra\RR/\ZZ$. Here ``Lagrangian'' means that the quadratic form vanishes when restricted to $\cL$, and any element orthogonal to $\cL$ (with respect to the bilinear form $\sb$ derived from the quadratic form) is in $\cL$. The Lagrangian subgroup $\cL$ describes the charges of Wilson lines which can meet at the surface operator.

\subsection{Algebra-objects corresponding to surface operators}
 
Now let us compute the algebra-object $W(S_v,\L)$ corresponding to a surface operator $S_v$ and a choice of a line operator $\L$ on which $S_v$ can terminate. Any object in the category of bulk line operators can be written in the form
$$
\bigoplus_{i =0}^{2N-1} V_i\otimes W_i,
$$
where the simple objects $W_i$ have been defined above. The ``expansion coefficients'' $V_i$ are finite-dimensional  complex vector spaces. To compute $V_i$ one needs to evaluate the vector space corresponding to $U(1)_{2N}$ Chern-Simons theory on $S^2\times\RR_t$ with an insertion of $W(S_v,\L)$ and the dual of $W_i$. Here $\RR_t$ is the time coordinate. An insertion of $W(S_v,\L)$ can be thought of as an insertion of a surface operator $S_v$ along a  submanifold $I\times\RR_t$, where $I$ is a segment of the meridian of $S^2$, see Fig. 13.

\begin{figure}[htbp] \label{fig19}
\begin{center}
\begin{picture}(4,4.2)(0,-0.2)
\put(1.6,3.8){$\L$}\put(1.6,-0.4){$\bar\L$}\put(3.7,2){$\overline W_i$}
\includegraphics{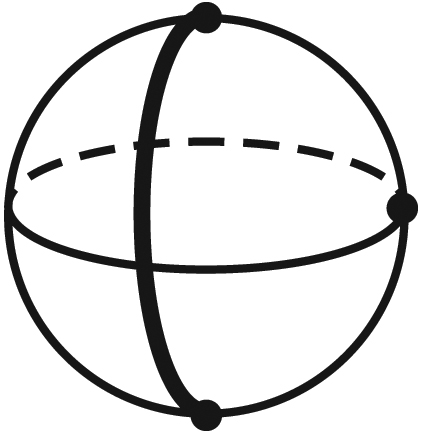}
\end{picture}
\end{center}

\caption{Computation of the algebra-object $W(S_v,\L)$. The surface operator $S_v$ is placed along a meridian of $S^2$ and terminates on line operators $\L$ and $\bar\L$. To get the ``expansion coefficient'' $V_i$ one also needs to insert the bulk Wilson line dual to $W_i$. }
\end{figure}

It is easy to see that the vector space $V_i$ is one-dimensional if the dual of $W_i$ can cancel the charge carried by the Wilson lines on which $S_v$ terminates, and is zero-dimensional otherwise. Thus all we need to determine is what bulk electric charge the surface operator bounded by $\L$ and $\bar\L$ can carry. It is convenient to use the folding trick: squash the $S^2$ in Fig. 13 along the meridian on which $S_v$ is placed. Then one gets $U(1)_{2N}\times U(1)_{-2N}$ Chern-Simons theory on a disk. The boundary of the disk is subdivided into two segments, and two different boundary conditions are imposed along these two segments. One of them is the boundary condition corresponding to the ``invisible'' surface operator, and the other one corresponds to the boundary condition $S_v$. At the two junctions of these segments one inserts line operators $\L$ and $\bar\L$. We would like to determine which bulk line operators can be inserted in the interior of the disk to cancel the charge created by $\L$ and $\bar\L$. Recall \cite{KS} that boundary conditions in such a theory correspond to Lagrangian subgroups in the finite abelian group $\cD\oplus \cD$ equipped with the quadratic form $\sq\oplus (-\sq)$, where elements of $\cD\oplus\cD\simeq \ZZ_{2N}\oplus\ZZ_{2N}$ label charges of bulk line operators. Th surface operator $S_v$ corresponds to the Lagrangian subgroup $\cL_v$ defined by (\ref{lagrv}). The ``invisible'' surface operator is obtained by setting $v=1$.

The meaning of this result is the following. The boundary gauge group is the subgroup of the bulk gauge group, and accordingly the group of boundary electric charges is the quotient of the group of bulk electric charges (i.e. it is a quotient of $\cD\oplus\cD$). The Lagrangian subgroup consists precisely of those bulk charges which project to zero in the group of boundary charges. Where boundary conditions corresponding to Lagrangian subgroups $\cL_1$ and $\cL_v$ meet along a line, the gauge group is broken further, and accordingly any bulk charge which belongs to $\cL_1\oplus\cL_v$ will be trivial as regards the unbroken gauge group at the junction. This has the following consequence: while the line operators sitting at the two ends of $I$ have opposite charges with respect to the gauge group left unbroken at the junction, the net charge may be an arbitrary element of $\cL_{tot}=\cL_1\oplus \cL_v$ when lifted to $\cD\oplus\cD$. 

The subgroup $\mathcal{L}_{tot}$ consists of elements of the form
$$(n_1+x,n_2-x),\quad n_1+n_2=0 \ \text{mod} \  2N/v,\quad  n_1-n_2=0 \  \text{mod} \  2v,\quad x=0,\ldots, 2N-1.$$
We want to cancel this charge by a Wilson line whose charge after the folding trick has the form $(Y,0)$, where $Y$ is an arbitrary element of $\cD$. 
Therefore we must require $$n_2=x,\quad Y=2Nj/v,$$ where $j$ is an arbitrary integer. Hence the algebra-object $W(S_v,\L)$ has the following expansion:
$$W(S_v,\L)=\bigoplus_{j=0}^{v-1}W_{jw},$$
where $w=2N/v$.
This is exactly the algebra-object denoted $\sA_w$ in \cite{FRS}. Thus we reproduced the classification of consistent RCFTs with a  $U(1)_{2N}$ current algebra.\footnote{In principle, we should also check that the Frobenius algebra structures agree. However, this is not really necessary since there is a unique (up to isomorphism) Frobenius algebra structure on the object $\AA_w$.} Note that all the symmetric Frobenius algebras $\sA_w$ happen to be special.

\subsection{Fusion of surface operators}

Surface operators in a 3d TFT can be fused together; this gives rise to a monoidal structure (i.e. a kind of tensor product) on the 2-category of surface operators. In general, fusing two surface operators which do not admit local operators may produce a surface operator which does admit nontrivial local operators. In this subsection we discuss fusion of surface operators in the $U(1)_{2N}$ Chern-Simons theory and compare with analogous results for algebra-objects in \cite{FRS}. A monoidal structure on a 2-category is a complicated notion which involves associator objects, 2-associators between tensor products of associator objects, and a coherence equation for 2-associators, see e.g. \cite{KV} for a detailed discussion. Here we will only discuss how the tensor product works on the level of objects.

As discussed above, a surface operator may be characterized by charges of Wilson lines which may meet at it. These charge live in a Lagrangian subgroup $\cL$ in $\cD\times\cD$. It is convenient to flip the orientation of the Wilson lines which comes in from the right, i.e. to think of it as an outgoing Wilson line instead of the incoming one. Thus instead of thinking about two Wilson lines coming to the surface from two different sides we will be thinking about a Wilson line which comes in from the left, pierces the surface operator and continues going in the same direction. A surface operator may cause a jump in the charge of a Wilson line from $X_1$ to $X_2$, and we will characterize a surface operator $S_v$ by the set of the allowed jumps. Clearly, this set is a subset of $\cD\times\cD$ which may be obtained by taking $\cL_v$ and flipping the sign of the second component. We will denote the resulting subset $R_v$ and will think of it as a relation between $\cD$ and $\cD$. 

The characterization of surface operators by relations is particularly useful when we consider the fusion of surface operators $S_v$ and $S_{v'}$. Clearly,  a Wilson line $W_a$ may turn into a Wilson line $W_c$ after meeting the composite surface operator $S_{v}\circ S_{v'}$ if and only if there exists $b\in\cD$ such that\footnote{If more than one such $b$ exists for a given pair $(a,c)$, then the composite surface operator is not merely $S_{v''}$ for some $v''$ but a sum of several such surface operators.} $(a,b)\in R_v$ and $(b,c)\in R_{v'}$.  In other words, the composition of surface operators corresponds to the composition of relations.

It easy to see that the surface operator $S_1$ is the identity with respect to the fusion product, as expected. Indeed, $S_1$ is determined by the Lagrangian subgroup $\cL_1$ consisting of elements of the form $(a,-a)$, $a\in\cD$, so the corresponding relation $R_1$ is the diagonal in $\cD\times\cD$, i.e. the identity relation. Another interesting case is $S_N$; according to \cite{FRS} the corresponding algebra-object $\sA_2$ is related to T-duality (i.e. the surface operator $S_N$ is the T-duality wall). Let us compute the fusion product of $S_v$ and $S_N$. The relation $R_N$ corresponding to $S_N$ is given by
$$
a+b=0,\quad a,b,\in\ZZ/2N\ZZ.
$$
Composing it with $R_v$ (in any order), we get a relation in $\cD\times\cD$ defined by pairs $(a,b)$ such that
$$
a-b=0 \ {\rm mod}\ 2v,\quad a+b=0\ {\rm mod}\ 2N/v.
$$
This is the relation $R_{N/v}$. Hence
\begin{equation}\label{Tdualitydefect}
S_v\circ S_N=S_N\circ S_v=S_{N/v}.
\end{equation}
In particular, we see that the surface operator $S_N$ is invertible: its inverse is $S_N$. This means that it implements a symmetry of the $U(1)_{2N}$ Chern-Simons theory. The relation (\ref{Tdualitydefect}) means that this symmetry maps the consistent gluing corresponding to $S_v$ to the consistent gluing corresponding to $S_{N/v}$. 
This agrees with \cite{FRS}.

In general, computing the composition of relations gives the following formula:
$$
S_v\circ S_{v'}=g S_{v''},
$$
where 
$$
g=\gcd(v,v',N/v,N/v'),\quad v''=\lcm(\gcd(v,N/v'),\gcd(v',N/v)).
$$
A proof (due to A. Tsymbalyuk) is given in the appendix. In particular we see that $S_v$ is invertible if and only if $\gcd(v,N/v)=1$, and that in this case the inverse is again $S_v$. These results agree with \cite{FRS}. 

\section{Boundary-bulk map in 3d TFT}

We have seen that to every surface operator $S$ and a line operator $\L$ on which it can terminate one can associate a symmetric Frobenius algebra in the ribbon category of bulk line operators. The proof made use of pictures which resemble open-string scattering diagrams. It is natural to ask if there is an analogous use for closed-string scattering diagrams in 3d TFT. In fact, closed-string scattering diagrams appear when one defines the 3d analogue of the boundary-bulk map familiar from 2d TFT. In the two-dimensional case for every object in the category of boundary conditions one can define a vector in the space of bulk local operators in the following way. Consider an annulus on whose interior boundary component one imposed the chosen boundary condition and whose exterior boundary component is a ``cut'' boundary. The path-integral of the 2d TFT on such a manifold defines a vector in the vector space associated to $S^1$, i.e. the space of local operators. From the categorical viewpoint, this space is the ``derived center'' (Hochschild cohomology) of the category of boundary conditions. This map from the set of boundary conditions to the space of bulk local operators is called the boundary-bulk map. 

If we consider the same annulus in a 3d TFT, it defines an object in the category associated to $S^1$. The latter category can be thought of either as the category of bulk line operators in the 3d TFT or as the Drinfeld center of the 2-category of boundary conditions. Thus we get a map from the set of boundary conditions to the set of bulk line operators. This is the 3d analogue of the boundary-bulk map.

It turns out that the bulk line operator one gets by applying the boundary-bulk map is very special: it is a commutative symmetric Frobenius algebra in the braided monoidal category of bulk line operators. This fact has no 2d analogue. To see why it is true one uses pictures resembling closed-string scattering diagrams. Let $\sA$ be the bulk line operator corresponding to some boundary condition. Pictorially, it is represented by a hollow cylindrical tube in $\RR^3$ on the boundary of which one imposes the boundary condition one is interested in. The product, coproduct, unit and counit arise from diagrams shown in Fig. 14. 

\begin{figure}[htbp] \label{fig20}
\begin{center}
\includegraphics{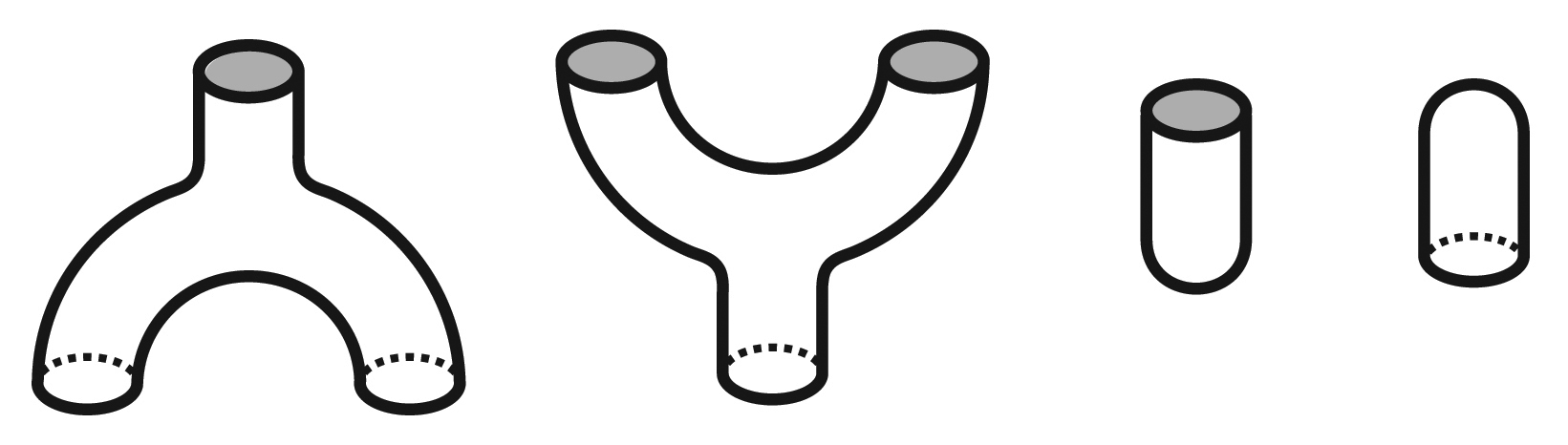}
\end{center}

\caption{Product, coproduct, unit and counit for the algebra $\AA$ corresponding to a boundary condition in a 3d TFT. The interior of the ``termite holes'' contains vacuum (trivial 3d TFT).}
\end{figure}

Commutativity means that the product morphism $m:\sA\otimes\sA\ra \sA$ is compatible with the self-braiding $c:\sA\otimes\sA\ra\sA\otimes\sA$:
$$
m\circ c=m.
$$
The algebra $\AA$ is commutative because the two diagrams shown in Fig. 15 can be deformed into one another.

\begin{figure}[htbp] \label{fig21}
\begin{center}
\includegraphics{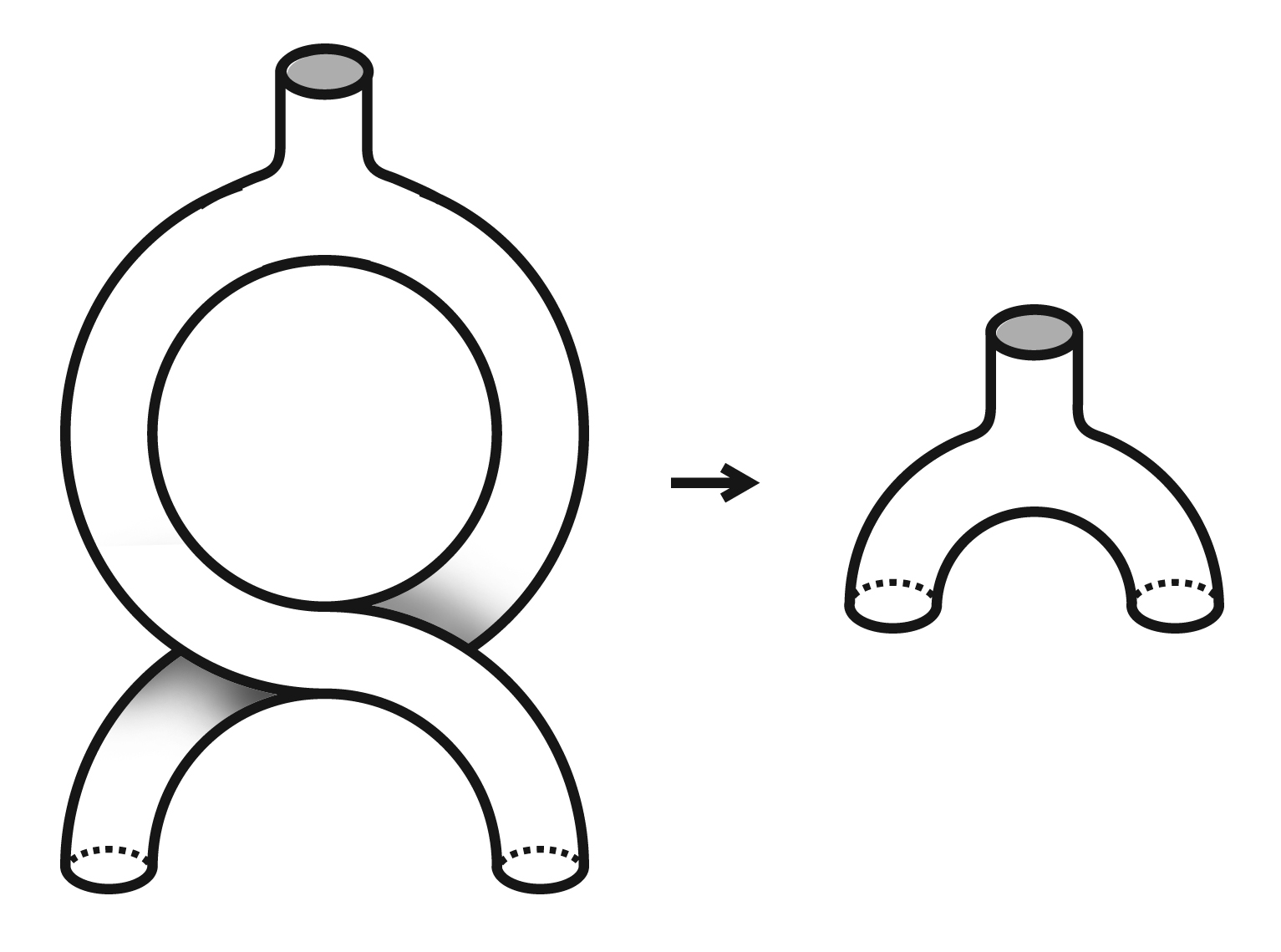}
\end{center}

\caption{The commutativity constraint.}
\end{figure}

To give a concrete example, consider $U(1)_{2N}\times U(1)_{-2N}$ Chern-Simons theory. Boundary conditions in this theory are the same as surface operators in $U(1)_{2N}$ theory and, as explained in the previous section, are classified by divisors of $N$. Alternatively, they are classified by subgroups of $\ZZ_{2N}\oplus\ZZ_{2N}$ which are Lagrangian with respect to the quadratic function
$$
\sq(X_1,X_2)=\frac{1}{4N}(X_1^2-X_2^2).
$$
If $v$ is a divisor of $N$, the corresponding Lagrangian subgroup $\cL_v$ is given by (\ref{lagrv}). We denote by $W(\cL_v)$ the bulk line operator corresponding to the Lagrangian subgroup $\cL_v$ via the 3d boundary-bulk map.

We would like to compute the expansion of $W(\cL_v)$ in terms of simple objects $W_X$ where $X\in \ZZ_{2N}\times \ZZ_{2N}$ labels the bulk electric charge. The ``expansion coefficient'' is a vector space $V_X$ which is the vector space corresponding to the disk with an insertion of a simple object ${\overline W}_X$ at the origin and with the boundary condition corresponding to $\cL_v$ imposed on the boundary. It is easy to see that $V_X$ is one-dimensional if $X$ has vanishing charge with respect to the boundary gauge group and is zero-dimensional otherwise. Since $\cL_v$ consists precisely of those bulk charges which are trivial with respect to the boundary gauge group, we conclude that $W(\cL_v)$ is given by\footnote{This is analogous to the following fact about Rozansky-Witten theory with target $\mathcal{X}$: the boundary-bulk map maps the boundary condition corresponding to a complex Lagrangian submanifold $\mathcal{Y}$ of $\mathcal{X}$ to the object of the derived category of $\mathcal{X}$ given by the push-forward of the structure sheaf of $\mathcal{Y}$ \cite{KRS}.}
$$
W(\cL_v)=\bigoplus_{X\in\cL_v} W_X.
$$
The product morphism $m:W(\cL_v)\otimes W(\cL_v)\ra W(\cL_v)$ is essentially determined up to isomorphism by charge conservation. Indeed, charge conservation tells us that $m$ maps the component $W_X\otimes W_Y=W_{X+Y}$ of $W(\cL_v)\otimes W(\cL_v)$ to the component $W_{X+Y}$ of $W(\cL_v)$. Assuming that this morphism is nonzero, we see that $m$ is determined by a $\CC^*$-valued function $m(X,Y)$ on $\cL_v\times\cL_v$. The associativity constraint reads
$$
(\delta m)(X,Y,Z)=a(X,Y,Z),\quad X,Y,Z\in\cL_v,
$$
where $\delta$ is the differential in the standard cochain complex computing cohomology of $\cL_v$ with coefficients in $\CC^*$, and $a(X,Y,Z)\in\CC^*$ is the restriction of the associator morphism to $\cL_v$. Since $\cL_v$ is isotropic (in fact, Lagrangian) with respect to the quadratic function $\sq$, the restriction of $a$ to $\cL_v$ is trivial \cite{KS}. This means that $m(X,Y)$ is a 2-cocycle. Further, recall that $W(\cL_v)$ is a commutative algebra. Since $\L_v$ is isotropic with respect to the quadratic function $\sq$, for any two $X,Y\in\cL_v$ the braiding is trivial. Hence the commutativity constraint becomes 
$$
m(X,Y)=m(Y,X).
$$
Such a 2-cocyle is necessarily cohomologically trivial \cite{Brown}, i.e. there exists a function $n:\cL_v\ra \CC^*$ such that 
$$
m(X,Y)=\frac{n(X+Y)}{n(X)n(Y)}.
$$
This means that $m(X,Y)$ can be made equal to $1$ by replacing the algebra $(W(\cL_v),m)$ with an isomorphic one, where the action of the isomorphism on the component $W_X$ of $W(\cL_v)$ is given by $n(X)$.

\section{Concluding remarks}

The main observation of this paper is that the classification of surface operators in a 3d TFT is equivalent to the classification of consistent gluings of chiral and anti-chiral sectors in the corresponding 2d RCFT. This observation led us to reinterpret the results of Fuchs, Runkel and Schweigert on the algebraic classification of consistent gluings in terms of surface operators. We considered a simple example of this relationship, $U(1)$ Chern-Simons theory, and showed that all 2d RCFTs with a $U(1)$ current algebra arise from surface operators. We note that the construction of gluing by means of surface operators seems to be more general than constructing of gluings by means of special symmetric Frobenius algebras. Indeed, while to any surface operator one can attach a symmetric Frobenius algebra (or rather, a Morita-equivalence class of symmetric Frobenius algebras), this algebra may fail to be special. In the case of $U(1)$ Chern-Simons theory all surface operators happen to yield special symmetric Frobenius algebras, so the two methods of constructing consistent gluings are equivalent.

Obviously, it would be desirable to classify surface operators in nonabelian Chern-Simons theory. However, the example of $SU(2)$ Chern-Simons theory indicates that this will not be easy. It is well known that 2d RCFTs with $SU(2)_k$ current algebra have ADE classification \cite{ade}. The A-type RCFTs exist for all levels $k$, the D-type RCFTs exist for all even $k$, while the E-type RCFTs exist for three ``sporadic'' values $k=10, 16,28$. This means that surface operators in $SU(2)$ Chern-Simons theory should have ADE classification as well. Clearly, the A-type surface operator must be the ``invisible'' surface operator. It is plausible that the D-type surface operator is defined by the following condition: if the $SU(2)$ gauge fields to the left and to the right of the surface operator are denoted $A^{(+)}$ and $A^{(-)}$, then the restrictions of $A^{(+)}$ and $A^{(-)}$ to the surface agree as $SO(3)$ gauge fields (but not necessarily as $SU(2)$ gauge fields). On the other hand, the E-type surface operators are much more mysterious, and the fact that they exist only for small enough $k$ suggests that they cannot be understood using semiclassical reasoning at all. 

It would also be interesting to extend our results to spin-topological 3d theories which depend on a choice of spin structure on a manifold.  The simplest example of such a theory is $U(1)$ Chern-Simons theory at an odd level $k$. This theory describes Fractional Quantum Hall phases and therefore a classification of codimension-1 defects in this theory would be of considerable interest.

\section*{Appendix}

\begin{prop} (A. Tsymbalyuk) 

Let $N$ be a natural number, $v$ be a divisor of $N$, and $R_v$ be a relation (in fact, a subgroup) in $\ZZ_{2N}\times\ZZ_{2N}$ given by
$$
a+b=0\ {\rm mod}\ 2v,\quad a-b=0\ {\rm mod}\ 2N/v.
$$
Then for any two divisors $v,v'$ we have
$$
R_v\circ R_{v'}=g R_x,
$$
where
$$
g=\gcd(v,v',N/v,N/v'),\quad x=\lcm(\gcd(v,N/v'),\gcd(v',N/v)).
$$

\end{prop}

\begin{proof}

Let $a,b,c$ be integers such that $(a,b)\in R_v$ and $(b,c)\in R_{v'}$. Since 
$$
a+c=(a-b)+(b+c)=(a+b)-(b-c),
$$
$a+c$ is divisible by $2 \gcd(v',N/v)$ as well as by $2 \gcd(v,N/v')$, and therefore is divisible by $2x$ where 
$$
x=\lcm(\gcd(v,N/v'),\gcd(v',N/v)).
$$
Similarly, since
$$
a-c=(a+b)-(b+c)=(a-b)+(b-c),
$$
$a-c$ is divisible by $2y$ where
$$
y=\lcm(\gcd(v,v'),\gcd(N/v,N/v').
$$
To prove that $R_v\circ R_{v'}\subset R_x$ it is sufficient to show that $xy=N$. Let $p$ be a prime factor of $N$, and let $\alpha=\ord_p N$, $i=\ord_p v$, $j=\ord_p v'$. Then 
$$
\ord_p x=\max(\min(i,\alpha-j),\min(\alpha-i,j)),\quad \ord_p y=\max(\min(i,j),\min(\alpha-i,\alpha-j)).
$$
Since there is a symmetry exchanging $v$ and $v'$ and $i$ and $j$, it is sufficient to consider the case $i\leq j$. Then there are four possible relative orderings of four numbers $i,j,\alpha-i,\alpha-j$. For each of them one can easily verify that $\ord_p x+\ord_p y=\ord_p N$. (For example, if $i\leq j\leq\alpha-j\leq \alpha-i$, then $\ord_p x=j$, $\ord_p y=\alpha-j$, and therefore $\ord_p x+\ord_p y=\alpha$). This proves that $xy=N$.

The opposite inclusion $R_v\circ R_{v'}\supset R_x$ follows from the Chinese remainder theorem. Therefore we have
$$
R_v\circ R_{v'}=g R_x
$$
for some natural number $g$.

Finally, let us compute $g$. Given a relation $R\subset \ZZ_{2N}\times\ZZ_{2N}$ we will say that $a\in\ZZ_{2N}$ is in the domain of $R$ if there exists $b\in\ZZ_{2N}$ such that 
$(a,b)\in R$. It follows from the Chinese remainder theorem that for a fixed $a$ in the domain of $R_v$ there exist $\gcd(v,N/v)$ solutions of the condition $(a,b')\in S_v$. Hence if $a$ is in the domain of $R_v\circ R_{v'}$, then there exist $\gcd(v,N/v)\cdot \gcd(v',N/v')$ solutions of the condition $(a,c')\in R_v\circ R_{v'}$. On the other hand, if $R_v\circ R_{v'}=g R_x$, this number must be equal to $g\cdot \gcd(x,N/x)$. Hence we have a formula for $g$:
$$
g=\frac{\gcd(v,N/v)\cdot \gcd(v',N/v')}{\gcd(x,y)}.
$$
The ratio on the right-hand-side of this formula is in fact equal to $\gcd(v,v',N/v,N/v')$. This is proved by analyzing the prime factorization of $N$ as above. Let $p$ be a prime factor 
of $N$, and $\alpha=\ord_p N$, $i=\ord_p v$, $\ord_p v'=j$. First of all, we have
$$
\ord_p g=\min(i,\alpha-i)+\min(j,\alpha-j)-\min(\ord_p x, \ord_p y).
$$
On the other hand, we have
$$
\ord_p (gcd(v,v',N/v,N/v'))=\min(i,j,\alpha-i,\alpha-j).
$$
Again we may assume without loss of generality that $i\leq j$, and then there are four cases to consider corresponding to four possible relative orderings of $i,j,\alpha-i,\alpha-j$. For example, if $i\leq j\leq\alpha-j\leq \alpha-i$, then $\ord_p g=i+j-j=i=\ord_p (gcd(v,v',N/v,N/v')).$ It is easy to verify that $\ord_p g=\ord_p (gcd(v,v',N/v,N/v'))$ in the other three cases as well. This concludes the proof.

\end{proof}

\end{document}